\documentclass[conference,twocolumn]{IEEEtran}
\pdfoutput=1

\newlength{\figwidth}
\usepackage{color}
\usepackage[nosort]{cite}        
\usepackage{url} 
\usepackage[intlimits]{amsmath}
\usepackage{bbm}
\usepackage{graphicx}
\usepackage{paralist}
\usepackage{fancyref}
\usepackage[stretch=16,shrink=16,step=4]{microtype}
\usepackage{vmr-symbols-vecbold}
\usepackage{standard-macros}

\makeatletter
\def\@IEEEinterspaceratioM{0.265}
\def\@IEEEinterspaceMINratioM{0.1651}
\def\@IEEEinterspaceMAXratioM{0.38}

\def\@IEEEinterspaceratioB{0.31}
\def\@IEEEinterspaceMINratioB{0.19}
\def\@IEEEinterspaceMAXratioB{0.38}
\@IEEEtunefonts
\makeatother
\newcommand{\genop}{\textnormal{gen}}
\newcommand{\gen}{\genop(W,Z^n)}
\newcommand{\infdensop}{\imath}
\newcommand{\infdens}{\infdensop(W,Z^n)}

\begin{document}

\IEEEoverridecommandlockouts

\title{Generalization Error Bounds via $m$th Central Moments of the Information Density}
% \author{Giuseppe~Durisi,~\IEEEmembership{Senior Member,~IEEE,}
%              %
% \thanks{}}%

%
\author{\IEEEauthorblockN{Fredrik Hellstr\"om, Giuseppe Durisi}\\
\IEEEauthorblockA{
Department of Electrical Engineering, Chalmers University of Technology, 41296 Gothenburg, Sweden\\
}}
\maketitle

%%%%%%%%%%%%%%%%
\begin{abstract}
We present a general approach to deriving bounds on the generalization error of randomized learning algorithms.
Our approach can be used to obtain bounds on the average generalization error as well as bounds on its tail probabilities, both for the case in which a new hypothesis is randomly generated every time the algorithm is used---as often assumed in the 
 probably approximately correct (PAC)-Bayesian
literature---and in the single-draw case, where the hypothesis is extracted only once.

For this last scenario, we present a novel bound that is explicit in the central moments of the information density.
The bound reveals that the higher the order of the information density moment that can be controlled, the milder the dependence of the generalization bound on the desired confidence level.

Furthermore, we use tools from binary hypothesis testing to derive a second bound, which is explicit in the tail of the information density. This bound confirms that a fast decay of the tail of the information density yields a more favorable dependence of the generalization bound on the confidence level.

\end{abstract}

%%%%%%%%%%%%%%%%%%%%%%%%%%%%%%%%%%
\section{Introduction}\label{sec:introduction}
A recent line of research, initiated by the work of Russo and Zou~\cite{russo16-05b} and then followed by many recent contributions~\cite{xu17-05a,bassily18-02a,bu19-01a,esposito19-12a}, has focused on obtaining bounds on the generalization error of randomized learning algorithms in terms of %which are explicit in
information-theoretic quantities, such as mutual information. 
The resulting bounds are \textit{deterministic}, i.e., data-independent, and allow one to assess the speed of convergence of a given learning algorithm in terms of sample complexity~\cite[p.~44]{shalev-shwartz14-a}.

A parallel development has taken place in the machine learning and statistics community, where the probably approximately correct (PAC)-Bayesian framework, pioneered by McAllester~\cite{mcallester98-07a}, has resulted in several upper bounds on the generalization error.
These bounds, which are expressed in terms of the relative entropy between a prior and a posterior %n \textit{a priori} and an \textit{a posteriori}
distribution on the hypothesis class (see, e.g.,~\cite{guedj19-01a} for a recent review), are
 typically \textit{empirical}, i.e., data-dependent, and can be used to design learning algorithms~\cite{catoni07-a}.

One difficulty in comparing the bounds on the generalization error available in the literature is that they sometimes pertain to different quantities.
To illustrate this point, we need to introduce some key quantities, which will be used in the remainder of the paper. 
Following the standard terminology in statistical learning theory, we let $\setZ$ be the instance space, \setW be the hypothesis space, and $\ell: \setW\times \setZ \rightarrow \positivereals$ be the loss function. 
A training data set $Z^n=[Z_1,\dots,Z_n]$  is a set of $n$ \iid samples drawn from a distribution $P_Z$ defined on $\setZ$.
We denote by $P_{Z^n}$ the product distribution induced by $P_Z$.
A randomized learning algorithm is characterized by a conditional probability distribution~$P_{W\!\given\! Z^n}$ on~$\mathcal{W}$.
Finally, we let the generalization error for a given hypothesis $w$ be defined as the difference between the population and empirical risks
\begin{equation}\label{eq:gen}
    \genop(w,z^n)=\frac{1}{n}\sum_{k=1}^{n}\ell(w,z_k) -\Ex{P_Z}{\ell(w,Z)}.
\end{equation}
Throughout the paper, we shall assume that the loss function $\ell(w,Z)$ is $\sigma$-subgaussian~\cite[Def.~2.2]{wainwright19-a} under $P_Z$ for all $w\in \setW$. 

The line of work initiated with~\cite{russo16-05b} deals with bounding the average generalization error
\begin{equation}\label{eq:average-gen}
    \Ex{P_{W\! Z^n}}{ \gen}.
\end{equation}
Specifically, upper bounds on the absolute value of this quantity were first presented in~\cite{russo16-05b} and then improved in~\cite[Thm.~1]{xu17-05a} and~\cite[Prop.~1]{bu19-01a}.

On the contrary, the PAC-Bayesian approach seeks lower bounds on the  probability~\cite{guedj19-01a}
\begin{equation}\label{eq:pac-bayesian}
  P_{Z^n}\lefto[\abs{\Ex{P_{W\!\given\! Z^n}}{\gen}} \leq \epsilon \right].
\end{equation}
Characterizing such a  probability, which is in the spirit of the PAC framework, is relevant when a new hypothesis $W$ is drawn from $P_{W\!\given\! Z^n}$ every time the algorithm is used.
As can be verified by, e.g., comparing the proof of~\cite[Lemma~1]{xu17-05a} and the proof of~\cite[Prop.~3]{guedj19-10a},\footnote{For the case in which the prior and posterior distributions in~\cite[Prop.~3]{guedj19-10a} are set to $P_W$ and $P_{W\!\given\! Z^n}$, respectively.} for the subgaussian case, one can obtain  bounds both on~\eqref{eq:average-gen} and on~\eqref{eq:pac-bayesian} that are explicit in the mutual information $I(W;Z^n)$ and in the relative entropy $\relent{P_{W\!\given\! Z^n}}{P_W}$, respectively, by using the Donsker-Varadhan variational formula for relative entropy.

One may also be interested in the scenario in which the hypothesis $W$ is drawn from $P_{W\!\given\! Z^n}$ only once, i.e., it is kept fixed for all uses of the algorithm.
In such a scenario, which, following the terminology used in~\cite[p.~12]{catoni07-a}, we shall refer to as a \textit{single-draw} scenario, the  probability of interest is 
\begin{equation}\label{eq:single-draw}
  P_{W\! Z^n}\lefto[\abs{\gen} \leq \epsilon \right].
\end{equation}
Bounds on this  probability that depend on the mutual information $I(W;Z^n)$ were provided in~\cite[Thm.~3]{xu17-05a} and~\cite{bassily18-02a}. 
Several novel bounds, which are explicit in information-theoretic quantities such as $f$-divergence, $\alpha$-mutual information, and maximal leakage, were recently derived in~\cite{esposito19-12a}.
Interestingly, all these bounds make use of a different set of tools compared with the ones used to establish bounds on~\eqref{eq:average-gen} and~\eqref{eq:pac-bayesian}, with one of the main ingredients being the data processing inequality for $f$-divergences.

Furthermore, they yield drastically different estimates for the generalization error. 
Specifically, let us assume that we want~\eqref{eq:single-draw} to be greater than $1-\delta$ where, throughout the paper, $\delta \in (0,1)$.
Then a slight refinement of the analysis in~\cite{bassily18-02a} yields the following bound on $\epsilon$:
\begin{equation}\label{eq:sample_complexity_mi}
    \epsilon\geq \sqrt{\frac{2\sigma^2}{n}\left(\frac{I(W;Z^n)+H_b(\delta)}{\delta}+\log 2\right)}. %
\end{equation}
Here, $H_b(\delta)$ denotes the binary entropy function. Throughout the paper, $\log(\cdot)$ denotes the natural logarithm.
In contrast, the analysis in~\cite[Cor.~5]{esposito19-12a}, yields the following bound for $\alpha>1$:
\begin{equation}\label{eq:sample_complexity_alpha_mi}
    \epsilon\geq \sqrt{\frac{2\sigma^2}{n} \left[I_{\alpha}(W;Z^n)+\log 2 + \frac{\alpha}{\alpha-1} \log \frac{1}{\delta}\right]}.
\end{equation}
Here, $I_{\alpha}(\cdot,\cdot)$ is the $\alpha$-mutual information
\begin{IEEEeqnarray}{rCl}\label{eq:alpha_MI}
I_\alpha(W;Z^n)& = &\frac{\alpha}{\alpha-1} \log \Exop_{P_{Z^n} }\lefto[
\Exop_{P_W}^{1/\alpha} \lefto[ \left(\frac{\dv P_{W\! Z^n}}{\dv P_W\! P_{Z^n}} \right)^\alpha \right] \right],\IEEEeqnarraynumspace
\end{IEEEeqnarray}
where $\dv P_{W\! Z^n}/\dv P_W\! P_{Z^n}$ is the Radon-Nikodym derivative. Note that, since $\lim_{\delta\to 0} H_b(\delta)/\delta+\log \delta = 1$, the dependence of $\epsilon$ on $\delta$ in~\eqref{eq:sample_complexity_mi} is of order $1/\sqrt{\delta}$.
In contrast, it is of order $\sqrt{(\alpha/(\alpha-1)) \log(1/\delta)}$ in~\eqref{eq:sample_complexity_alpha_mi}, which is typically more favorable.
For example, in the limit $\alpha\to\infty$, the $\alpha$-mutual information converges to the maximal leakage~\cite[Thm.~1]{issa16-a}, and $\epsilon$ depends on $\delta$ only through the term $\sqrt{\log(1/\delta)}$.

The analysis in~\cite{esposito19-12a}, however, does not reveal why using $\alpha$-mutual information rather than mutual information results in a more benign dependence of the generalization error on the confidence parameter $\delta$.
Moreover, the choice $\alpha=1$, for which $I_{\alpha}(W;Z^n)$ reduces to $I(W;Z^n)$, renders the bound in~\eqref{eq:sample_complexity_alpha_mi} vacuous.

\paragraph*{Contributions}
Inspired by the treatment of the generalization error for the case of the $0-1$ loss function reported in~\cite{catoni07-a}, we present a single framework for deriving bounds on the generalization error that can be applied to both average and tail analyses, both of a  PAC-Bayesian and single-draw flavor.
As a product 
of our analysis, we obtain a probabilistic generalization error bound for the single-draw scenario, which results in the following bound on $\epsilon$ to guarantee that~\eqref{eq:single-draw} is greater than $1-\delta$:
\begin{equation}\label{eq:sample_complexity_moments}
    \epsilon\geq  \sqrt{\frac{2\sigma^2}{n}\left(I(W;Z^n)+\frac{M_m(W;Z^n)}{(\delta/2)^{1/m}}+\log\frac{2}{\delta}\right)}.
\end{equation}
Here,
\begin{equation}\label{eq:central_moment_infodens}
    M_m(W;Z^n) = \Exop^{1/m}_{P_{W\! Z^n}}\lefto[\abs{ \imath(W,Z^n) - I(W;Z^n)}^m \right]
\end{equation}
is the $m$th root of the 
$m$th central moment of the information density
\begin{equation}\label{eq:info_density}
    \imath(w,z^n)=\log \frac{\dv P_{W\! Z^n} }{\dv P_W\! P_{Z^n}}(w,z^n).
\end{equation}
The bound in~\eqref{eq:sample_complexity_moments} is derived as a data-independent relaxation of an underlying data-dependent bound. Comparing~\eqref{eq:sample_complexity_mi} with~\eqref{eq:sample_complexity_moments}, we see that the existence of higher central moments of $\imath(W,Z^n)$ results in a more favorable scaling of the error bound with~$\delta$.
This implies that one can obtain generalization error bounds that are explicit in the mutual information and have a more favorable dependence on $\delta$ than the one given in~\eqref{eq:sample_complexity_mi}.
In the limit $m\to\infty$, the dependence is of order $\sqrt{\log (1/\delta)}$, but the resulting bound is less tight than the maximal leakage bound in~\cite[Cor.~5]{esposito19-12a}. However, through a more refined analysis, we recover the maximal leakage bound, up to a logarithmic term.

To shed further light on the role of the tail of the information density in determining the dependence of $\epsilon$ on $\delta$, we derive an additional probabilistic single-draw bound, based on a change of measure argument~\cite[Thm.~12.5]{polyanskiy19-a} that is used to establish strong converse bounds in binary hypothesis testing. 
It results in the following bound on $\epsilon$:
\begin{IEEEeqnarray}{rCl}
    \epsilon&\geq&\sqrt{ \frac{2\sigma^2}{n}
    \left(
        \gamma + \log\lefto(\frac{2}{\delta-P_{W\! Z^n}\lefto[\imath(W,Z^n)\geq \gamma\right]} \right)
    \right)}.\IEEEeqnarraynumspace\label{eq:strongconv_gen_thm_intro}
\end{IEEEeqnarray}
Similar to~\eqref{eq:sample_complexity_moments}, this bound reveals that for a fixed~$\delta$, low values of~$\epsilon$ require fast-decaying tails of the information density random variable.
Indeed, $\gamma$ in~\eqref{eq:strongconv_gen_thm_intro} should be chosen sufficiently large to make the argument of the $\log$ positive. 
However, large values of~$\gamma$ also contribute to a large $\epsilon$.

\section{Bounds via a Subgaussian Inequality}\label{sec:main_results}
 In this section, we derive several types of bounds on the absolute value of the generalization error of a randomized learning algorithm.
The following theorem gives an inequality that will later be used to derive both average and tail bounds for the generalization error.
\begin{thm}\label{thm:mainineq}
Let  $Z^n$ be \iid according to $P_Z$. Assume that $\ell(w,Z)$ is $\sigma$-subgaussian under $P_Z$ for all $w\in \mathcal W$. Assume that $P_{W\! Z^n}$ is absolutely continuous with respect to $P_W\! P_{Z^n}$. Then, for all $\lambda\in \reals$,
\begin{IEEEeqnarray}{rCl}\label{eq:mainineq}
    \Ex{P_{W\! Z^n}}{\exp\lefto(\lambda \gen-\frac{\lambda^2\sigma^2}{2n}-\imath(W,Z^n)  \right)}&\leq& 1.\IEEEeqnarraynumspace
\end{IEEEeqnarray}
\end{thm}
\begin{IEEEproof}
Since $\ell(w,Z)$ is $\sigma$-subgaussian and the $Z_i$ are \iid, the random variable $\frac{1}{n}\sum_{i=1}^n  \ell(w,Z_i)$ is $\sigma/\sqrt{n}$-subgaussian, i.e.,
\begin{multline}
    \Exop_{P_{Z^n} } \lefto[\exp\lefto(\lambda\left(\frac{1}{n}\sum_{i=1}^n \ell(w,Z_i) - \Exop_{P_Z}\lefto[\ell(w,Z)\right] \right)\right)\right] \\
    \leq \exp\lefto(\frac{\lambda^2\sigma^2}{2n}\right).
\end{multline}
Reorganizing terms and taking the expectation with respect to $P_W$, we get
\begin{equation}\label{eq:subgauss_without_indicator}
    \Exop_{P_W\! P_{Z^n} } \lefto[\exp\lefto(\lambda \gen- \frac{\lambda^2\sigma^2}{2n}\right)\right]\leq 1.
\end{equation}
Now, let $E$ be the union of all sets $\setE\in \mathcal{W}\times \mathcal{Z}^n$ such that $P_{W\!Z^n}(\setE)=0$, and let $\bar E$ denote its complement. It follows from~\eqref{eq:subgauss_without_indicator} that
\begin{equation}
    \Exop_{P_W\! P_{Z^n} } \lefto[1_{\bar E}\cdot\exp\lefto(\lambda \gen- \frac{\lambda^2\sigma^2}{2n}\right)\right]\leq 1,
\end{equation}
where $1_{\bar E}$ is the indicator function of the set $\bar E$. To obtain~\eqref{eq:mainineq}, we perform a change of measure from $P_W\! P_{Z^n}$  to $P_{W\! Z^n}$, as per~\cite[Prop.~17.1(4)]{polyanskiy19-a}.
\end{IEEEproof}

We next show how the inequality~\eqref{eq:mainineq} can be used to derive previously known and novel bounds on the generalization error.

\subsection{Average Generalization Error}
As a first corollary of Theorem~\ref{thm:mainineq}, we derive a bound on the average generalization error~\eqref{eq:average-gen}, recovering the result in~\cite[Thm.~1]{xu17-05a}.% without an explicit use of the Donsker-Varadhan variational formula for relative entropy.
\begin{cor}\label{cor:ExpectedGen}
Under the assumptions of Theorem~\ref{thm:mainineq},
\begin{equation}
    \abs{ \Ex{P_{W\! Z^n}}{\gen}} \leq \sqrt{\frac{2\sigma^2}{n}I(W;Z^n)}.
\end{equation}
\end{cor}
\begin{IEEEproof}
We apply Jensen's inequality to~\eqref{eq:mainineq}, which yields
\begin{equation}\label{eq:intermediate_average}
    \exp\biggl(\lambda\Ex{P_{W\! Z^n}}{\gen} - \frac{\lambda^2\sigma^2}{2n} 
 -\Ex{P_{W\! Z^n}}{\imath(W,Z^n)} \biggr)  \leq   1.
\end{equation}
% \begin{IEEEeqnarray}{rrCl}\label{eq:intermediate_average}
% \exp\biggl(\lambda\Ex{P_{W\! Z^n}}{\gen} - \frac{\lambda^2\sigma^2}{2n} & & \nonumber\\
% -\Ex{P_{W\! Z^n}}{\imath(W,Z^n} &\biggr) & \leq  & 1.
% \end{IEEEeqnarray}
Noting that $\Ex{P_{W\! Z^n}}{\imath(W,Z^n)}=I(W;Z^n)$ we get, after taking the $\log$ of both sides of~\eqref{eq:intermediate_average} and reorganizing terms, the nonnegative parabola in $\lambda$
\begin{equation}
    \lambda^2\frac{\sigma^2}{2n} - \lambda \Ex{P_{W\! Z^n}}{\gen} + I(W;Z^n) \geq 0.
\end{equation}
Since the discriminant of a nonnegative parabola is nonpositive, we get
%Its discriminant  must be nonpositive. Therefore,
\begin{equation}
\Exop^2_{P_{W\! Z^n}}[\gen] - \frac{2\sigma^2}{n}I(W;Z^n)\leq 0,
\end{equation}
which yields the desired bound.
\end{IEEEproof}

\subsection{PAC-Bayesian Tail Bounds}\label{sec:pac-bayesian-results}
Next, we use Theorem~\ref{thm:mainineq} to obtain two tail bounds on the absolute value of the generalization error averaged over $P_{W\!\given\! Z^n}$ in~\eqref{eq:pac-bayesian}.
The first one, presented in Corollary~\ref{cor:PAC-bayesian}, recovers a classical data-dependent PAC-Bayesian bound (see, e.g.,~\cite[Prop.~3]{guedj19-10a}) for the special case in which $P_W$ is taken as the prior distribution and $P_{W\!\given\! Z^n}$ is taken as the posterior distribution.
The second one, presented in Corollary~\ref{cor:PAC-bayesian-data-independent}, is a relaxation of the first bound, which makes it data-independent.
This bound, which depends on the $m$th moment of the relative entropy $\relent{P_{W\vert Z^n}}{P_W}$, recovers the bound given in~\cite[App.~A.3]{bassily18-02a} for the case $m=1$.
\begin{cor}\label{cor:PAC-bayesian}
Under the assumptions in Theorem~\ref{thm:mainineq}, the following bound holds with probability at least $1-\delta$ under $P_{Z^n}$:
\begin{multline}
    \abs{ \Exop_{P_{W\vert Z^n} }\gen } 
     \\ \leq \sqrt{\frac{2\sigma^2}{n}\left(\relent{P_{W\vert Z^n}}{P_W} +\log \frac{1}{\delta}\right)}.\label{eq:multidrawBase}
     \end{multline}
%      \begin{multline}
%     \abs{ \Exop_{P_{W\vert Z^n} } \gen } 
%     \\ \leq \sqrt{\frac{2\sigma^2}{n}\left(\frac
%     {\Exop^{1/m}_{P_{Z^n}}\lefto[\relent{P_{W\vert Z^n}}{P_W}^m\right] }
%     {(2/\delta)^{1/m}} +\log \frac{2}{\delta}\right)}.\label{eq:multidrawMI}
% \end{multline}
\end{cor}
\begin{IEEEproof}
Similarly to the proof of Corollary~\ref{cor:ExpectedGen}, we apply Jensen's inequality to~\eqref{eq:mainineq}, but now only with respect to the conditional expectation of $W$ given $Z^n$.
This yields
\begin{IEEEeqnarray}{rrCl}
\Exop_{P_{Z^n}} \biggo[\exp\biggo(\lambda \Ex{P_{W\vert Z^n} }{\gen} - \frac{\lambda^2\sigma^2}{2n} & & & \nonumber \\
-\relent{P_{W\!\given\! Z^n}}{P_W}
\bigg) &\bigg] &\leq& 1,
\label{eq:singleDrawJensen}\end{IEEEeqnarray}
where we used that \begin{equation}\Ex{P_{W\vert Z^n=z^n}}{i(W,z^n)}=\relent{P_{W\vert Z^n=z^n}}{P_W}.\end{equation}
%\begin{equation}\Ex{P_{W\!\given\! Z^n}}{i(W,Z^n)}=\relent{P_{W\!\given\! Z^n}}{P_W}.\end{equation}
Next, we use Markov's inequality in the following form: let $U\distas P_U$ be a nonnegative random variable s.t. $\Ex{}{U}\leq 1$.
Then
\begin{align}
P_U[U>1/\delta]< \Ex{}{U}\delta\leq \delta.\label{eq:MarkovTrick}
\end{align}
Using~\eqref{eq:MarkovTrick} in~\eqref{eq:singleDrawJensen}, we conclude that
\begin{IEEEeqnarray}{rrCl}
P_{Z^n}%\lefto
\biggo[\exp%\lefto
\biggo(\lambda \Ex{P_{W\vert Z^n} } {\gen} - \frac{\lambda^2\sigma^2}{2n}
 & & & \nonumber\\-\relent{P_{W\!\given\! Z^n}}{P_W} %\right
\bigg) \leq \frac{1}{\delta} %\right
&\bigg] &\geq & 1-\delta. \IEEEeqnarraynumspace
\end{IEEEeqnarray}
Reorganizing terms, we obtain:
\begin{IEEEeqnarray}{rrCl}
    P_{Z^n}\biggo[\frac{\lambda^2\sigma^2}{2n} -\lambda \Ex{P_{W\vert Z^n} } {\gen}  & & & \nonumber\\
    +\relent{P_{W\!\given\! Z^n}}{P_W} +\log\frac{1}{\delta} \geq 0 
&\bigg] &\geq 1&-\delta.
\end{IEEEeqnarray}
The desired bound~\eqref{eq:multidrawBase} now follows from the same discriminant analysis as in the proof of  Corollary~\ref{cor:ExpectedGen}.
\end{IEEEproof}

The bound in Corollary~\ref{cor:PAC-bayesian} is data-dependent because the upper bound on the generalization error depends on the specific instance of $Z^n$.
In the next corollary, we apply Markov's inequality once more to make the bound data-independent.

\begin{cor}\label{cor:PAC-bayesian-data-independent}
Under the assumptions in Theorem~\ref{thm:mainineq}, the following bound holds with probability at least $1-\delta$ under $P_{Z^n}$ for all $m>0$:
  \begin{multline}
    \abs{ \Ex{P_{W\vert Z^n} } {\gen} } 
    \\ \leq \sqrt{\frac{2\sigma^2}{n}\left(\frac
    {\Exop^{1/m}_{P_{Z^n}}\lefto[\relent{P_{W\vert Z^n}}{P_W}^m\right] }
    {(\delta/2)^{1/m}} +\log \frac{2}{\delta}\right)}.\label{eq:multidrawMI}
\end{multline}
\end{cor}

\begin{IEEEproof}
Applying Markov's inequality to the random variable $\relent{P_{W\vert Z^n}}{P_W}^m$, we obtain after some manipulations
\begin{multline}
P_{Z^n}\Biggl[\relent{P_{W\vert Z^n}}{P_W} \leq 
\frac
    {\Exop^{1/m}_{P_{Z^n}}\lefto[\relent{P_{W\vert Z^n}}{P_W}^m\right] }
    {\delta^{1/m}} \Biggr]\\
    \geq 1-\delta. \label{eq:MultidrawProofMarkov}
\end{multline}
We now observe that the two probability bounds~\eqref{eq:multidrawBase} and~\eqref{eq:MultidrawProofMarkov} together with the union bound imply that, with probability at least $1-2\delta$ under $P_{Z^n}$,
\begin{multline}
    \abs{ \Ex{P_{W\vert Z^n} } {\gen} }\\ \leq  \sqrt{\frac{2\sigma^2}{n}\left(\frac
    {\Exop^{1/m}_{P_{Z^n}}\lefto[\relent{P_{W\vert Z^n}}{P_W}^m\right] }
    {\delta^{1/m}} +\log \frac{1}{\delta}\right)}.
\end{multline}
The desired result then follows by the substitution $\delta\rightarrow \delta/2$. 
\end{IEEEproof}

Note that when $m=1$, we have 
\begin{equation}
    \Exop_{P_{Z^n}}\lefto[\relent{P_{W\vert Z^n}}{P_W}\right]=I(W;Z^n)
\end{equation}
and the bound~\eqref{eq:multidrawMI} coincides with the one reported in~\cite[App.~3]{bassily18-02a}.
Some additional remarks on~\eqref{eq:multidrawMI} are provided in Section~\ref{sec:remarks}. 

\subsection{Single-Draw Probabilistic Bounds}\label{sec:sigle-draw-results}
We now use Theorem~\ref{thm:mainineq} to derive tail bounds on the absolute value of the single-draw generalization error in~\eqref{eq:single-draw}.
As in Section~\ref{sec:pac-bayesian-results}, we first state a data-dependent bound in Corollary~\ref{cor:single-draw-data-dependent}.
Then, we relax this to two different data-independent bounds in Corollaries~\ref{cor:single-draw-data-independent} and~\ref{cor:single-draw-maximal-leakish}.
To the best of our knowledge, the first two bounds are novel, while the third recovers~\cite[Cor.~10]{esposito19-12a} up to a logarithmic term.
\begin{cor}\label{cor:single-draw-data-dependent}
Under the assumptions in Theorem~\ref{thm:mainineq}, the following bound holds with probability at least $1-\delta$ under $P_{W\! Z^n}$:\footnote{Note that the argument of the square root can be negative, but that this happens with probability at most $\delta$. Therefore, the right-hand side of~\eqref{eq:singledrawBase} is well-defined with probability at least $1-\delta$.}
\begin{equation}\label{eq:singledrawBase}
    \abs{ \gen } \leq \sqrt{\frac{2\sigma^2}{n}\left(\imath(W,Z^n) +\log \frac{1}{\delta}\right)}.
    \end{equation}
\end{cor}

\begin{IEEEproof}
Applying Markov's inequality~\eqref{eq:MarkovTrick} directly to~\eqref{eq:mainineq}, we conclude that
\begin{multline}
P_{W\! Z^n}\biggo[\exp\biggo(\lambda \gen - \frac{\lambda^2\sigma^2 }{2n}
-\infdens \bigg)\leq \frac{1}{\delta} \bigg]\\ \geq 1-\delta,
\end{multline}
from which the desired result follows by the same discriminant analysis as in the proof of Corollary~\ref{cor:ExpectedGen}.
\end{IEEEproof}

\begin{cor}\label{cor:single-draw-data-independent}
Under the assumptions in Theorem~\ref{thm:mainineq}, the following bound holds with probability at least $1-\delta$ under $P_{W\! Z^n}$:
  \begin{multline}
    \abs{  \gen } \leq \label{eq:singledrawMoment} \\
     \sqrt{\frac{2\sigma^2}{n}\left(I(W;Z^n)+\frac{M_m(W;Z^n)}{(\delta/2)^{1/m}} +\log \frac{2}{\delta}\right)},
\end{multline}
where $M_m(W;Z^n)$, defined in~\eqref{eq:central_moment_infodens}, is the $m$th root of the %added this
$m$th central moment of the information density. 
\end{cor}
\begin{IEEEproof}
We shall use Markov's inequality in the following form: 
for a random variable $U$,
\begin{equation}\label{eq:markov-for-arbitrary-rv}
%U \leq \abs{U-\Exop U} + \Exop U \leq \frac{\Exop^{1/m}\abs{U-\Exop U}^m}{\delta^{1/m}} + \Exop U.
P_U\lefto[U \leq \Exop[U]+ \frac{\Exop^{1/m}[\abs{U-\Exop[U]}^m]}{\delta^{1/m}} \right]\geq 1-\delta.
\end{equation}
Applying~\eqref{eq:markov-for-arbitrary-rv} to the information density random variable, we conclude that, with probability at least $1-\delta$,
\begin{equation}\label{eq:moment_bound_on_inf_dens}
    \infdens \leq I(W;Z^n)+\frac{M_m(W;Z^n)}{\delta^{1/m}}.
\end{equation}
It now follows from~\eqref{eq:singledrawBase},~\eqref{eq:moment_bound_on_inf_dens}, and the union bound that, with probability at least $1-2\delta$, 
\begin{multline}
\abs{\gen}\leq \\
\sqrt{\frac{2\sigma^2}{n}\left(I(W;Z^n)+\frac{M_m(W;Z^n)}{\delta^{1/m}}+\log \frac{1}{\delta}   \right)}.
\end{multline}
The desired result follows after the substitution $\delta\rightarrow \delta/2$.
\end{IEEEproof} 
\begin{cor}\label{cor:single-draw-maximal-leakish}
Under the assumptions in Theorem~\ref{thm:mainineq}, the following bound holds with probability at least $1-\delta$ under $P_{W\! Z^n}$:
  \begin{equation}
    \abs{  \gen } \leq \label{eq:singledrawMaximalLeakish}
     \sqrt{\frac{2\sigma^2}{n}\left(\mathcal{L}(Z^n \rightarrow W) +2\log \frac{2}{\delta}\right)}.
  \end{equation}
Here, $\mathcal{L}(Z^n \rightarrow W)$ denotes the maximal leakage, defined as
\begin{equation}
\mathcal{L}(Z^n \rightarrow W) =  \log \Ex{P_W}{\esssup_{P_{Z^n}} \frac{\dv P_{W\! Z^n}}{\dv P_W\!P_{Z^n}}}.
\end{equation}
\end{cor}
\begin{IEEEproof}
Markov's inequality implies that, with probability at least $1-\delta$ under $P_{W\!Z^n}$,
\begin{equation}\label{eq:cor_7_proof_step_1}
\infdens \leq \log \Ex{P_{W\!Z^n}}{\frac{\dv P_{W\! Z^n}}{\dv P_W\!P_{Z^n}}} + \log\lefto(\frac{1}{\delta}\right)
\end{equation}
Next, we can bound the expectation over~$P_{Z^n\vert W}$ by an essential supremum:
\begin{align}
\Ex{P_W\!P_{Z^n\vert W}}{\frac{\dv P_{W\! Z^n}}{\dv P_W\!P_{Z^n}}} \leq \Ex{P_W}{\esssup_{P_{Z^n\vert W}} \frac{\dv P_{W\! Z^n}}{\dv P_W\!P_{Z^n}}}.
\end{align}
The assumption that $P_{W\!Z^n}\ll P_W\!P_{Z^n}$ means that any set in the support of $P_{W\!Z^n}$ is also in the support of $P_W\!P_{Z^n}$. We can therefore upper-bound the $\esssup$ as follows:
\begin{align}\label{eq:cor_7_proof_step_3}
\esssup_{P_{Z^n\vert W}} \frac{\dv P_{W\! Z^n}}{\dv P_W\!P_{Z^n}} \leq \esssup_{P_{Z^n}} \frac{\dv P_{W\! Z^n}}{\dv P_W\!P_{Z^n}}.
\end{align}
Combining~\eqref{eq:cor_7_proof_step_1}-\eqref{eq:cor_7_proof_step_3}, we see that
\begin{equation}
\infdens \leq \log \mathcal{L}(Z^n \rightarrow W) + \log\lefto(\frac{1}{\delta}\right),
\end{equation}
which, combined with~\eqref{eq:singledrawBase} through the union bound and the substitution $\delta\rightarrow \delta/2$, gives the desired result.
\end{IEEEproof}
\subsection{Remarks on the Tail Bounds in Sections~\ref{sec:pac-bayesian-results} and~\ref{sec:sigle-draw-results}}\label{sec:remarks}
The single-draw tail bound in~\eqref{eq:singledrawMoment} reveals a relation between the central moments of the information density and the confidence parameter $\delta$. 
Specifically, the higher the moment of the information density that can be controlled, the more benign the dependence of the generalization error bound on $\delta$.
A similar observation holds for the data-independent PAC-Bayesian bound~\eqref{eq:multidrawMI}, in which  controlling higher moments of the random variable $\relent{P_{W\!\given\! Z^n}}{P_W}$ leads to a more favorable dependence of the generalization bound on $\delta$. 

In the limit $m\to\infty$ the bound in~\eqref{eq:singledrawMoment} reduces to 
 \begin{multline}\label{eq:singledrawMomentInf}
    \abs{  \gen } \leq  \\
     \sqrt{\frac{2\sigma^2}{n}\left(I(W;Z^n)+{M_\infty(W;Z^n)} +\log \frac{2}{\delta}\right)},
\end{multline}
where $M_\infty(W;Z^n)=\esssup_{P_{W\!Z^n}} \abs{\imath(w,z^n)- I(W;Z^n)}$.
%So, if $M_\infty(W,Z^n)$ is finite, the dependence on $\delta$ is of order $\log(1/\delta)$.
So, in this limit, the dependence on $\delta$ is of order $\sqrt{\log(1/\delta)}$.
However, the bound~\eqref{eq:singledrawMaximalLeakish} is tighter than~\eqref{eq:singledrawMomentInf}, up to the factor~$2$ multiplying the logarithm. It is also tighter than the max information bound in~\cite[Thm.~4]{dwork15-06a} with $\beta=0$, up to the aforementioned factor of~$2$. Indeed, let the max information be defined as
\begin{equation}\label{eq:max_mi}
    I_{\textnormal{max}}(W;Z^n) = \esssup_{P_{W\!Z^n} }\imath(w,z^n).
\end{equation}
It is readily verified that
\begin{equation}I_\textnormal{max}(W;Z^n)\leq I(W;Z^n)+{M_\infty(W;Z^n)}.
\end{equation}
As shown in~\cite[Lem.~12]{esposito19-12a}, $\mathcal{L}(Z^n \rightarrow W)\leq I_{\textnormal{max}}(W;Z^n)$. Thus, provided that
\begin{equation}\mathcal{L}(Z^n \rightarrow W) \leq I_\textnormal{max}(W;Z^n) + \log\frac{2}{\delta},
\end{equation}
we have established that the bound in~\eqref{eq:singledrawMaximalLeakish} is stronger than, in order, the max information bound in~\cite[Thm.~4]{dwork15-06a} with $\beta=0$, and~\eqref{eq:singledrawMomentInf}. However, the maximal leakage bound in~\cite[Cor.~10]{esposito19-12a} is still stronger than the one in~\eqref{eq:singledrawMaximalLeakish} by a~$\log 2/\delta$ term inside the square root.

In the next section, we present a different approach to obtaining single-draw tail bounds, which reveals a coupling between~$\delta$ and the tail of the information density random variable. 

\section{Bounds via the Strong Converse}
As pointed out in Section~\ref{sec:introduction}, a key tool for deriving the single-draw bound~\eqref{eq:sample_complexity_mi} is the data processing inequality for $f$-divergences. This is also true for some %Changed many -> some. It only really applies to the alpha-divergence and f-divergence bounds, and the former is a special case of the latter
of the bounds presented  in~\cite{esposito19-12a}.
In the context of binary hypothesis testing, it is known that such an inequality only leads to a weak converse bound on the region of achievable error rates.
To obtain a strong converse, one needs to use~\cite[Lem.~12.2]{polyanskiy19-a} (restated in Lemma~\ref{lem:strong_converse_lemma} below for convenience), which provides a bound on the probability of an event under a distribution $P$ in terms of its probability under~$Q$.
\begin{lem}\label{lem:strong_converse_lemma}
Let $E$ be an arbitrary event and $P$ and $Q$ be probability measures such that $P$ is absolutely continuous with respect to $Q$.
Then, for all $\gamma\in\reals$,
\begin{equation}\label{lem:StrongConv}
    P[E] \leq P\lefto[\log \frac{\dv P}{\dv Q} >  \gamma \right] + e^\gamma Q[E].
\end{equation}
\end{lem}
As we shall show next,  this inequality can be turned into a generalization bound by choosing $P$, $Q$, and $E$ appropriately. 
\begin{thm}\label{thm:strong_conv_bound}
Under the assumptions of Theorem~\ref{thm:mainineq}, the following bound holds with probability at least $1-\delta$ over $P_{W\! Z^n}$:
\begin{multline}\label{eq:strongconv_gen_thm}
    \abs{\gen}\\
    \leq\sqrt{ \frac{2\sigma^2}{n}
    \left(
        \gamma + \log\lefto(\frac{2}{\delta-P_{W\! Z^n}\lefto[\imath(W,Z^n)\geq \gamma\right]} \right)
    \right)}
\end{multline}
for all $\gamma$ for which the arguments of the logarithm and the square root are nonnegative.
\end{thm}
\begin{IEEEproof}
With $P=P_{W\! Z^n}$, $Q=P_W\! P_{Z^n}$ and 
\begin{equation}\label{eq:high_error_event}
E=\{(w,z^n): \abs{\textnormal{gen}(w,z^n)} > \epsilon  \},
\end{equation}
we apply Lemma~\ref{lem:strong_converse_lemma} to get
\begin{equation}\label{eq:StrongConvPf1}
P_{W\! Z^n} [E] \leq {P_{W\! Z^n} }\left[\imath(W,Z^n)\geq \gamma  \right] + e^\gamma {P_W\! P_{Z^n}}[E].
\end{equation}
The $\sigma$-subgaussianity of the loss function implies that~\cite[Eq.~(2.9)]{wainwright19-a}
\begin{equation}
{P_{Z^n}}\lefto[\abs{\textnormal{gen}(w,Z^n)}>\epsilon\right] \leq 2\exp\lefto(-n{\epsilon^2}/{(2\sigma^2)} \right).\label{eq:hoeffding_q_bound}
\end{equation}
Inserting~\eqref{eq:hoeffding_q_bound} into \eqref{eq:StrongConvPf1}, we obtain
\begin{multline}\label{eq:strong_conv_bound}
    P_{W\! Z^n}[\abs{\gen}>\epsilon] \\\leq P_{W\! Z^n}\lefto[\imath(W,Z^n)\geq \gamma\right] + 2\exp\lefto(\gamma-n{\epsilon^2}/{(2\sigma^2)}\right).
\end{multline}
We get the desired result by imposing that the right-hand side of~\eqref{eq:strong_conv_bound} is less than $\delta$ and solving for~$\epsilon$.
\end{IEEEproof}

Unlike the bounds in Section~\ref{sec:main_results}, this bound depends on the tail distribution of the information density. 
For a given~$\delta$, the parameter $\gamma$ needs to be chosen large enough to make the factor~$\delta-P_{W\! Z^n}[\imath(W,Z^n)\geq \gamma]$ positive.
However, choosing~$\gamma$ too large makes the bound loose because of the $\gamma$ term that is added to the~$\log$. This reveals a trade-off between the rate of decay of the tail of the information density and the confidence level~$\delta$.

Controlling the tail of the information density results in a tighter bound than the moment-based bound in~\eqref{eq:singledrawMoment} and the maximal leakage bound~
\cite[Cor.~10]{esposito19-12a} (up to some~$\log 1/\delta$ terms). Indeed, these two bounds can be obtained  by further upper-bounding the right-hand side of~\eqref{eq:strongconv_gen_thm}, as we shall discuss next.

\subsection{Moment-Based Single-Draw Tail Bound}\label{sec:rederive_sd_tail_moments}
By Markov's inequality,
\begin{IEEEeqnarray}{rCl}
& &P_{W\! Z^n}\lefto[\imath(W,Z^n)\geq \gamma\right]\notag \\%& =&P_{W\! Z^n}\lefto[\imath(W,Z^n)\geq \gamma-I(W;Z^n) + I(W;Z^n)\right] \\%&\leq & P_{W\! Z^n}\lefto[I(W;Z^n)+\abs{\imath(W,Z^n)-I(W;Z^n)}\geq \gamma' + I(W;Z^n)\right]\nonumber\\
&\leq& P_{W\! Z^n}\lefto[\abs{\imath(W,Z^n)-I(W;Z^n)}\geq \gamma-I(W;Z^n)\right] \IEEEeqnarraynumspace\\
&\leq & \frac{(M_m(W;Z^n))^m }{(\gamma-I(W;Z^n))^{m}}.\label{eq:strong_conv_rederive_2}
\end{IEEEeqnarray}
We now set
\begin{align}\label{eq:gamma}
    \gamma = \frac{M_m(W;Z^n)}{(\delta/2)^{1/m}}+I(W;Z^n).
\end{align}
Subsituting~\eqref{eq:gamma} in~\eqref{eq:strong_conv_rederive_2}, we conclude that 
\begin{equation}
    P_{W\! Z^n}\lefto[\imath(W,Z^n)\geq \gamma\right] \leq {\delta}/{2}.
\end{equation}
%\begin{IEEEeqnarray}{rClCl}& & P_{W\! Z^n}\lefto[\imath(W,Z^n)\geq \gamma\right] &\leq& \frac{(M_m(W;Z^n))^m }{\left(\frac{M_m(W;Z^n)}{(\delta/2)^{1/m}}\right)^{m}}\\& & & = & \frac{\delta}{2}.\end{IEEEeqnarray}
Inserting this upper bound into \eqref{eq:strongconv_gen_thm} we obtain
\begin{equation}
\epsilon \leq \sqrt{\frac{2\sigma^2}{n} \left(I(W;Z^n)+\frac{M_m(W;Z^n)}{(\delta/2)^{1/m}} + \log\frac{4}{\delta} \right)},
\end{equation}
 which coincides with \eqref{eq:singledrawMoment}, up to a $\log 2$ term. 

\subsection{Maximal Leakage Single-Draw Tail Bound}\label{sec:rederive_max_info}
Using the assumption that $P_{W\!Z^n}\ll P_W\!P_{Z^n}$, we get
\begin{equation}
P_{W\! Z^n}[\infdens \geq \gamma] \leq P_W\lefto[\esssup_{P_{Z^n}}\frac{\dv P_{W\! Z^n}}{\dv P_W\!P_{Z^n}}\geq e^\gamma\right].
\end{equation}
Thus, Markov's inequality implies that
\begin{equation}
P_{W\! Z^n}[\infdens \geq \gamma] \leq e^{-\gamma}\Ex{P_W}{\esssup_{P_{Z^n}}\frac{\dv P_{W\! Z^n}}{\dv P_W\!P_{Z^n}}}.
\end{equation}
Setting $\gamma = \mathcal{L}(Z^n \rightarrow W) + \log(2/\delta)$ and using this result in~\eqref{eq:strongconv_gen_thm}, we get, with probability at least $1-\delta$ over $P_{W\! Z^n}$,
\begin{IEEEeqnarray}{rCl}
\abs{\gen}&\leq& \sqrt{ \frac{2\sigma^2}{n}\left(\mathcal{L}(Z^n\rightarrow W)+\log \frac{4}{\delta}+\log \frac{2}{\delta} \right) }.\IEEEeqnarraynumspace
\end{IEEEeqnarray}

%%%%%%%%%%%%%%%%%%%%%%%%%%%%
\bibliographystyle{IEEEtran}
\bibliography{reference}

\begin{thebibliography}{10}
\providecommand{\url}[1]{#1}
\csname url@samestyle\endcsname
\providecommand{\newblock}{\relax}
\providecommand{\bibinfo}[2]{#2}
\providecommand{\BIBentrySTDinterwordspacing}{\spaceskip=0pt\relax}
\providecommand{\BIBentryALTinterwordstretchfactor}{4}
\providecommand{\BIBentryALTinterwordspacing}{\spaceskip=\fontdimen2\font plus
\BIBentryALTinterwordstretchfactor\fontdimen3\font minus
  \fontdimen4\font\relax}
\providecommand{\BIBforeignlanguage}[2]{{%
\expandafter\ifx\csname l@#1\endcsname\relax
\typeout{** WARNING: IEEEtran.bst: No hyphenation pattern has been}%
\typeout{** loaded for the language `#1'. Using the pattern for}%
\typeout{** the default language instead.}%
\else
\language=\csname l@#1\endcsname
\fi
#2}}
\providecommand{\BIBdecl}{\relax}
\BIBdecl

\bibitem{russo16-05b}
D.~Russo and J.~Zou, ``\BIBforeignlanguage{en}{Controlling {Bias} in {Adaptive}
  {Data} {Analysis} {Using} {Information} {Theory}},'' in
  \emph{\BIBforeignlanguage{en}{Artificial {Intelligence} and {Statistics}}},
  May 2016, pp. 1232--1240.

\bibitem{xu17-05a}
A.~Xu and M.~Raginsky, ``Information-theoretic analysis of generalization
  capability of learning algorithms,'' in \emph{Advances in Neural Information
  Processing Systems}, 2017, pp. 2524--2533.

\bibitem{bassily18-02a}
R.~Bassily, S.~Moran, I.~Nachum, J.~Shafer, and A.~Yehudayoff, ``Learners that
  {{Use Little Information}},'' \emph{Proc. Algorithmic Learning Theory, PLMR},
  vol.~83, no. 25-55, 2018.

\bibitem{bu19-01a}
Y.~Bu, S.~Zou, and V.~V. Veeravalli, ``Tightening {Mutual} {Information}
  {Based} {Bounds} on {Generalization} {Error},'' Jan. 2019, arXiv: 1901.04609.

\bibitem{esposito19-12a}
\BIBentryALTinterwordspacing
A.~R. Esposito, M.~Gastpar, and I.~Issa, ``Generalization error bounds via
  {R{\`e}nyi} $f$-divergences and maximal leakage,'' Dec. 2019, arXiv.
  [Online]. Available: \url{http://arxiv.org/abs/1912.01439}
\BIBentrySTDinterwordspacing

\bibitem{shalev-shwartz14-a}
S.~Shalev-Shwartz and S.~Ben-David, \emph{Understanding machine learning: from
  theory to algorithms}.\hskip 1em plus 0.5em minus 0.4em\relax Cambridge,
  U.K.: Cambridge Univ. Press, 2014.

\bibitem{mcallester98-07a}
D.~A. McAllester, ``Some {PAC}-bayesian theorems,'' in \emph{Proc. Conf.
  Computational Learning Theory (COLT)}, Jul. 1998, pp. 230--234.

\bibitem{guedj19-01a}
\BIBentryALTinterwordspacing
B.~Guedj, ``A {Primer} on {PAC}-{Bayesian} {Learning},'' Jan. 2019, arXiv.
  [Online]. Available: \url{http://arxiv.org/abs/1901.05353}
\BIBentrySTDinterwordspacing

\bibitem{catoni07-a}
O.~Catoni, ``{PAC}-{{Bayesian Supervised Classification}}: {{The
  Thermodynamics}} of {{Statistical Learning}},'' \emph{IMS Lecture Notes
  Monogr. Ser.}, vol.~56, pp. 1--163, 2007.

\bibitem{wainwright19-a}
M.~J. Wainwright, \emph{High-dimensional statistics: a nonasymptotic
  viewpoint}.\hskip 1em plus 0.5em minus 0.4em\relax Cambridge, U.K.: Cambridge
  Univ. Press, 2019.

\bibitem{guedj19-10a}
\BIBentryALTinterwordspacing
B.~Guedj and L.~Pujol, ``Still no free lunches: the price to pay for tighter
  {PAC}-{Bayes} bounds,'' Oct. 2019, arXiv. [Online]. Available:
  \url{http://arxiv.org/abs/1910.04460}
\BIBentrySTDinterwordspacing

\bibitem{issa16-a}
I.~{Issa}, S.~{Kamath}, and A.~B. {Wagner}, ``An operational measure of
  information leakage,'' in \emph{2016 Annual Conference on Information Science
  and Systems (CISS)}, March 2016, pp. 234--239.

\bibitem{polyanskiy19-a}
Y.~Polyanskiy and Y.~Wu, \emph{Lecture Notes On Information Theory}, Cambridge,
  U.K., 2019.

\bibitem{dwork15-06a}
C.~Dwork, V.~Feldman, M.~Hardt, T.~Pitassi, O.~Reingold, and A.~Roth,
  ``Generalization in adaptive data analysis and holdout reuse,'' in
  \emph{Advances in Neural Information Processing Systems}, 2015, pp.
  2350--2358.

\end{thebibliography}
\end{document}